\documentclass[lettersize,journal]{IEEEtran}
\usepackage{amsmath,amsfonts}
\usepackage{algorithmic}
\usepackage{algorithm}
\usepackage{array}
\usepackage[caption=false,font=normalsize,labelfont=sf,textfont=sf]{subfig}
\usepackage{textcomp}
\usepackage{stfloats}
\usepackage{url}
\usepackage{verbatim}
\usepackage{graphicx}
\usepackage{tikz}
\usepackage{cite}
\usepackage{titlesec}
\usepackage{xcolor}
\usepackage{pifont}

\newcommand{\revise}[1]{\textcolor{black}{#1}}

\titlespacing{\section}{0pt}{*0.7}{*0.7}
\titlespacing{\subsection}{0pt}{*0.5}{*0.5}

\begin{document}

\title{Latent-Feature-Informed Neural ODE Modeling for Lightweight Stability Evaluation of Black-box Grid-Tied Inverters }



\author{Jialin Zheng,~\IEEEmembership{Member,~IEEE,}
Zhong Liu,~\IEEEmembership{Student Member,~IEEE,}
Xiaonan Lu,~\IEEEmembership{Member,~IEEE}
  \vspace{-30pt}
}

\markboth{Journal of \LaTeX\ Class Files,~Vol.~14, No.~8, August~2025}%
{Shell \MakeLowercase{\textit{et al.}}: A Sample Article Using IEEEtran.cls for IEEE Journals}


\maketitle

\begin{abstract}
Stability evaluation of black-box grid-tied inverters is vital for grid reliability, yet identification techniques are both data-hungry and blocked by proprietary internals. \revise{To solve this, this letter proposes a latent-feature-informed neural ordinary differential equation (LFI-NODE) modeling method that can achieve lightweight stability evaluation directly from trajectory data.} LFI-NODE parameterizes the entire system ODE with a single continuous-time neural network, allowing each new sample to refine a unified global model. It faithfully captures nonlinear large-signal dynamics to preserve uniform predictive accuracy as the inverter transitions between operating points. Meanwhile, latent perturbation features distilled from every trajectory steer the learning process and concurrently reveal the small-signal eigenstructure essential for rigorous stability analysis. Validated on a grid-forming inverter, \revise{The LFI-NODE requires one to two orders of magnitude fewer training samples compared with traditional methods, collected from short time-domain trajectories instead of extensive frequency-domain measurements.} \revise{Furthermore, the LFI-NODE requires only 48 short transients to achieve a trajectory prediction error at the hundredth level and an eigenvalue estimation error at the tenth level, outperforming benchmark methods by one to two orders of magnitude.} This makes LFI-NODE a practical and lightweight approach for achieving high-fidelity stability assessment of complex black-box power-electronic systems.

\end{abstract}

\vspace{-5pt}
\vspace{-5pt}
\section{Introduction}
\IEEEPARstart{G}{rid}-tied inverters (GTIs) are essential interfaces connecting renewable energy resources to electrical grids. The rapid proliferation of converters and increasingly diverse control strategies (e.g., Grid-Forming [GFM] and Grid-Following [GFL]) have amplified complex dynamic interactions, posing significant challenges on system stability such as wide-frequency-range small-signal oscillations \cite{ref4}. Consequently, rapid and accurate stability assessments of GTIs are critical for the reliable operation of modern power systems.

Stability evaluation methods based on control theory are well-developed. The most widely used approach is state-space eigenvalue analysis, a technique that has been applied extensively in laboratory prototypes and custom-built test beds, with numerous refined variants now available \cite{ref5, 4118327}. While these methods are physically insightful, they rely on white-box scenarios defined by the system's explicit governing ordinary differential equations (ODEs), which require detailed inverter information that manufacturers often do not disclose as proprietary \cite{ref8}. This makes stability assessment for the vast number of commercial black-box inverters exceedingly difficult.

For such black-box scenarios, impedance-based stability criteria such as Nyquist methods are viable alternatives, since impedance can be derived from measured inverter responses to external perturbations \cite{ref6}. However, impedance models are derived through linearization at specific operating points and are therefore only valid under those conditions. Any changes in the system's operating conditions necessitate a new measurement and modeling cycle \cite{ref7}. To ensure model validity across a global operating range, extensive perturbation tests are required, which is both expensive and time-consuming.

To avoid repeating impedance test,  data-driven methods have emerged that utilize neural networks (NNs) to create surrogate models for black-box converters. Current research in this area can be divided into two main categories. The first category involves constructing time-domain models using discrete networks such as Recurrent Neural Networks (RNNs) or Nonlinear Autoregressive with Exogenous Inputs (NARX) network \cite{ref2,ref3}. However, these models are typically trained with trajectory fidelity, making them naturally emphasize large-signal dynamics. Besides, their discrete point-to-point mapping structures struggle to represent the intrinsic system dynamics, extracting the subtle small-signal dynamics required for stability analysis becomes exceedingly difficult.  

In contrast, the second category involves frequency-domain NNs thatdirectly learn the mapping to system impedance \cite{ref8}.  \revise{ While these approaches are more direct for impedance-based analysis, it is fundamentally rooted in learning from a large collection of local and linearized models. The significant variance among these local models imposes a data-intensive path to achieving robust generalization. Even recent advanced methods that demonstrate impressive generalization by using stacked-autoencoder, they still have not escaped the reliance on massive data, typically requiring over a thousand impedance profiles measured with specialized frequency-sweep equipment \cite{10748372}.} Crucially, both approaches face a fundamental barrier: GTIs are not inherently data-rich systems. The difficulty in acquiring sufficient perturbation data to train any high-fidelity model severely impedes their practical implementation.

To address these challenges,  this letter propsoes a lightweight latent-feature-informed neural ODE (LFI-NODE) modeling method tailored for black-box GTI stability assessment. Departing from traditional methods that separately model local large- and small-signal dynamics, the proposed method aims to obtain a unified intrinsic model that captures both regimes. To do this, LFI-NODE uses a single continuous NN structure to parameterize the system's governing ODE, so all data reinforce one global model instead of being scattered across local surrogates. Full details of the trajectories are learned using a high-precision numerical solver, and latent disturbance features extracted from those trajectories are added as informed guidance terms. This combined strategy empowers the model to learn the true governing ODEs, allowing it to replicate both large- and small-signal dynamics with high fidelity from a \revise{data-efficiency} dataset, thereby realizing its lightweight objective.


The main contributions can be summarized as follows:

\begin{itemize}
    \item A unified modeling paradigm that directly learns the intrinsic ODEs of black-box inverters, overcoming the data-inefficiency of traditional methods.
    \item A continuous-time Neural network for learning the GTI's complete dynamics, enabling a global model to capture different nonlinear large-signal dynamics.
    \item A latent-feature-informed training mechanism, where stability-critical information extracted from system responses is embedded as a regularizing signal to ensure the model's fidelity in small-signal dynamics.
\end{itemize}


\section{Unified Intrinsic Modeling Paradigm}

\begin{figure*}[t]
\begin{minipage}{\textwidth}
\centering
\vspace{-0.4cm}
\includegraphics[width=0.9\textwidth]{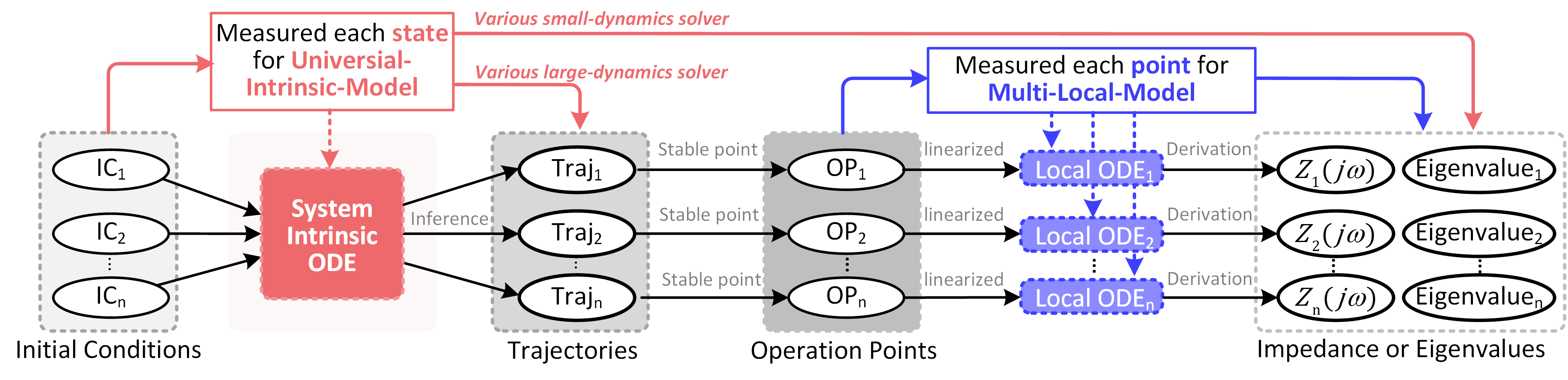}
\vspace{-0.4cm}
\caption{The comparison between the universal-intrinsic-modeling paradigm (red) and multi-local-modeling paradigm (blue).}
\label{Ref_UniverisalODE}
\end{minipage}
\vspace{-0.6cm}
\end{figure*}

\subsection{Structural Limitations of Multi-Local-Model Paradigm}

Figure \ref{Ref_UniverisalODE} illustrates the complete stability evaluation workflow for a dynamical system, from its initial conditions to the calculation of impedance or eigenvalues at operating points. In contrast, conventional data-driven methods for small-signal stability assessment are based on a mapping from the system's steady-state operating point to its measured impedance, as depicted by the blue section in Fig. \ref{Ref_UniverisalODE}.  This process begins by linearizing the system's dynamics at a given operating point, resulting in a local linear model represented by: 
\begin{equation} \label{ref_ode_func}
\Delta \dot{x} = A \Delta x + B \Delta u 
\end{equation}
Subsequently, a Fourier or Laplace transform is applied to this local model to derive the frequency-domain impedance $Z(j\omega; \eta)$. 
The critical issue is that the model training does not learn the intrinsic small-signal dynamics represented by the matrices $A$ and $B$. Instead, it implicitly learns a composite mapping that connects an operating point to the frequency-domain solution of its corresponding linearized local model.

The primary limitation of this 'multi-local model' paradigm is that it yields a collection of independent local ODEs with no transferable relationship between them. This structural limitation leads to poor generalization, making it difficult to predict behavior at new operating points. Consequently, achieving acceptable accuracy across a wide range requires the costly and inefficient process of exhaustively collecting data from numerous operating points.

\subsection{Unified-Intrinsic-Modeling Paradigm}
To overcome these limitations, we propose a unified-intrinsic-modeling paradigm shift from learning multiple local models to identifying the single intrinsic ODE that governs the system's complete dynamics (the red path in Fig. \ref{Ref_UniverisalODE}). This is achieved by directly learning the continuous vector field from data, yielding a unified surrogate model.

This single governing ODE offers major advantages: 1) it enables high-fidelity time-domain simulation via numerical solvers; 2) it can be analytically linearized at any operating point for small-signal analysis (e.g., impedance, eigenvalues), maintaining full compatibility with conventional tools.  \revise{Crucially, this unified modeling paradigm is inherently data-efficient, as its data requirement scales with the system's state dimension rather than the number of operating points to be covered, thus resolving the challenge of data scarcity in practical applications.} This provides a true end-to-end model suitable for a full spectrum of stability studies.

To unlock these benefits, the learning architecture itself must produce an explicit continuous-time vector field. However, conventional discrete-time networks such as RNNs and LSTMs are ill-suited: they approximate step-to-step state transitions rather than the underlying differential dynamics, so they cannot be analytically linearized to obtain the state matrices $(A,B)$ required for standard small-signal stability assessments. A dedicated continuous-time architecture is therefore essential for intrinsic ODE learning, \revise{a choice motivated by its direct compatibility with eigenvalue-based stability analysis and its sufficiency for modeling the component-level dynamics of a single inverter.}

\section{Latent-Feature-Informed Neural ODE}

\subsection{Neural Parameterization of Inverter Dynamics}
\begin{figure*}[t]
\begin{minipage}{\textwidth}
\centering
\vspace{-0.5cm}
\includegraphics[width=0.9\textwidth]{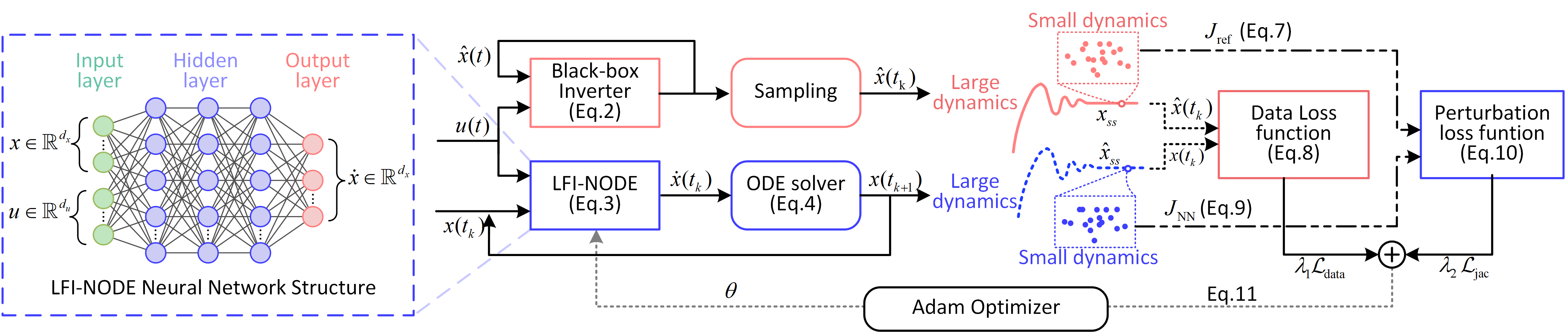}
\vspace{-0.3cm}
\caption{Schematic diagram of the neural network structure and learning process of the LFI-NODE approach.}
\label{Ref_LFI-NODE_diagram}
\end{minipage}
\vspace{-0.7cm}
\end{figure*}

Electrical transient dynamics of power inverters inherently constitute continuous-time processes, and Eq. \ref{ref_ode_func} can be expanded into the following parametric form, as shown in left side of Fig. \ref{Ref_LFI-NODE_diagram}:
\begin{equation} 
\dot{x} = f(x, u; \theta)
\end{equation}
where $x \in \mathbb{R}^{d_x}$ is  $d_x$-dimensional system-state vector, $u \in \mathbb{R}^{d_u}$ is a $d_u$-dimensional external input vector, and $\theta$  represents the intrinsic parameters that define the system's dynamics.

Due to unknown parameters $\theta$ and highly nonlinear system ODE functions $f$, we parameterize the function using the following neural networks:
\begin{equation} 
f_{\mathrm{NN}}(z)
= W_n\,\sigma\bigl(\cdots\sigma(W_1 z + b_1)\cdots\bigr) + b_n
\end{equation}
where $\sigma$ is the activation function (e.g., ReLU or tanh), $W_i \in \mathbb{R}^{d_{h_i} \times d_{h_{i-1}}}$, $\quad b_i \in \mathbb{R}^{d_{h_i}}$ are the learnable weight matrix and bias vector of layer $i$, respectively.  The input $z$ of the network is spliced by the state vector $x$ and the control vector $u$, i.e., $z = \left[x^T,  u^T \right]^T$, enabling the network to learn the system's dynamic response to various inputs.

Given an initial state $x(t_0)$, the system's state trajectory can be predicted by numerically integrating the learned neural ODE function $f _{\mathrm{NN}} $ using a numerical ODE solver:
\begin{equation}
\mathbf{x}_{k+1} = \mathbf{x}_k + \text{ODESolver}(f_{\mathrm{NN}}, \mathbf{x}_k, \mathbf{u}_k, \Delta t_k)
\end{equation}

Directly parameterizing ODE equations allows for learning continuous functions that explicitly represent system dynamics, providing a finer-grained and smoother dynamic approximation. This approach achieves higher accuracy compared to direct mapping methods such as RNN/NARX, which are limited to learning error compensations at fixed time steps, making it more suitable for the ODE-based framework.

\subsection{Latent Perturbation Feature Extraction}
Although Neural ODEs (NODEs) are proficient at capturing large-signal dynamics, their finite capacity can limit their ability to accurately model the small-signal behavior essential for stability analysis. The learned function $f_{\mathrm{NN}}$ may not precisely reflect the system's local dynamics around an equilibrium point. To address this, we introduce a mechanism to extract these small-signal characteristics directly from observational data, thereby making the system's implicit local stability features explicit and available to guide the training process.

The extraction procedure begins by identifying the steady-state equilibrium point $x_{\mathrm{ss}} $ , from a given dynamic trajectory $x(t) $. This is achieved by finding the time interval where the norm of the state's derivative  $\dot x(t) $, is minimal and taking the mean of the state vector over this period. In the vicinity of this equilibrium, the nonlinear dynamics $\dot x = f(x) $ can be linearized via a first-order Taylor expansion as:
\begin{equation}
\Delta\dot{x} \approx J\,\Delta x,
\end{equation}
where $\Delta x = x - x_{\mathrm{ss}} $ represents the state deviation and  $J $ is the system's Jacobian matrix at $x_{\mathrm{ss}}$ . To estimate $J $ numerically , we select  $N $ neighboring samples $\left\{ x_{i} \right\}_{i=1}^{N}
 $ around the equilibrium and compute their derivatives $ \dot x_i $ via first-order finite differences on the measured trajectory. The resulting deviation–derivative pairs $(x_i, \dot{x}_i) $ are stacked column-wise into the matrices,
\begin{align}
\Delta X = [\Delta x_1, \dots, \Delta x_N],  \quad
 \Delta \dot{X} = [\dot{x}_1, \dots, \dot{x}_N].
\end{align}

The estimation of $J_{\mathrm{ref}}$  is thus framed as a least-squares regression problem. A numerically robust solution is obtained using the Moore-Penrose pseudoinverse:
\begin{equation} \label{ref_Jacobii}
  J_{\mathrm{ref}} = \Delta\dot{X}\,\mathrm{pinv}(\Delta X)
\end{equation}
This data-driven matrix $J_{\mathrm{ref}} $ serves as a compact and potent latent feature that encapsulates the target system's small-signal characteristics, which can be used to inform the NN learning.

\revise{We acknowledge that, in practical applications, estimating the Jacobian via numerical differentiation can be sensitive to measurement noise. To improve robustness, we apply two measures. First, before performing numerical differentiation, the measured state trajectories are processed with a zero-phase low-pass filter \cite{9852391}, which effectively suppresses high-frequency noise without introducing additional phase shift. Second, as shown in (\ref{ref_Jacobii}), the Jacobian is estimated through a least-squares fit using up to $N$ sampled points. This approach inherently averages noise across multiple measurements, making it more robust than methods that rely on only a few points. A detailed mathematical analysis of noise effects, including an upper bound on the estimation error, is provided in Appendix.}

\subsection{Latent-Feature-Informed Learning Mechanism}

To ensure the Neural ODE model uniformly represents both large- and small-signal dynamics, we designed an end-to-end learning mechanism guided by the latent Jacobian feature, as is shown in Fig. \ref{Ref_LFI-NODE_diagram}. The training process optimizes the model's parameters $\theta$ by minimizing a composite loss function that simultaneously constrains the model's adherence to the system's global trajectory and its local stability characteristics.

The first component of this loss function $\mathcal{L}_{\mathrm{data}}$  ensures large-signal accuracy by minimizing the discrepancy between the predicted trajectory  $\hat{x}(t) $ and a measured reference trajectory $x_{\mathrm{ref}}(t) $. This is typically quantified using the mean squared error (MSE):
\begin{equation}
   \mathcal{L}_{\mathrm{data}}
= \frac{1}{T} \sum_{k=1}^{T} \bigl\|\hat{x}(t_k) - x(t_k)\bigr\|_2^2
\end{equation}

The second component $\mathcal{L}_{\mathrm{jac}}$ enforces small-signal fidelity. It measures the difference between the reference Jacobian $J_{\mathrm{ref}}$ and the model's Jacobian $J_{\mathrm{NN}}$. $J_{\mathrm{ref}}$ can be extracted from data using Eq. \ref{ref_Jacobii}, and $J_{\mathrm{NN}}$ can be  computed precisely at the predicted equilibrium point  $\hat{x}_{\mathrm{ss}} $ using automatic differentiation:
\begin{equation}
   J_{\mathrm{NN}}
   = \left.\frac{\partial f_{\mathrm{NN}}(x,u)}{\partial x}\right|_{x=\hat{x}_{\mathrm{ss}}}
\end{equation}
The discrepancy is then quantified by the squared Frobenius norm of their difference:
\begin{equation}
     \mathcal{L}_{\mathrm{jac}}
   = \bigl\|J_{\mathrm{NN}} - J_{\mathrm{ref}}\bigr\|_F^2
\end{equation}

The overall training objective is to minimize the total loss, which is a weighted sum of these two components:
\begin{equation}
      \mathcal{L}_{\text{total}} = \lambda_1 \,\mathcal{L}_{\mathrm{data}}+ \lambda_2 \,\mathcal{L}_{\mathrm{jac}}
\end{equation}

\revise{This composite loss function contains the trajectory-fitting term $\mathcal{L}_{\mathrm{data}}$ and the Jacobian-matching term $\mathcal{L}_{\mathrm{jac}}$. It is the core of our method. The design changes the learning task from simple trajectory fitting to a multi-objective optimization with physical constraints. The hyperparameters $\lambda_1$ and $\lambda_2$ control the trade-off between reproducing large-signal trajectories and matching small-signal stability.}

\revise{Specifically, $\mathcal{L}_{\mathrm{data}}$ ensures the accurancy of large-signal behavior. It minimizes the difference between predicted and measured trajectories so that the model can capture the full nonlinear dynamic behavior under disturbances. Meanwhile, the term $\mathcal{L}_{\mathrm{jac}}$ ensures  the accurancy of small-signal behavior. It acts as a physics-informed regularizer and forces the model’s linearized behavior near equilibrium points to match the true system. This introduces first-order derivative information into the training and constrains the solution space. By combining both terms, the LFI-NODE framework learns the same underlying ODE from two complementary views. The trajectory data describe the global dynamic path, and the Jacobian data describe the local tangent space. This dual constraint links large-signal fitting with small-signal stability analysis, improves generalization, and avoids convergence to solutions without physical meaning.}


\section{Case Study: Full-Order GFM Inverter}
\begin{figure}[t] 
    \centering
    \includegraphics[width=0.44\textwidth]{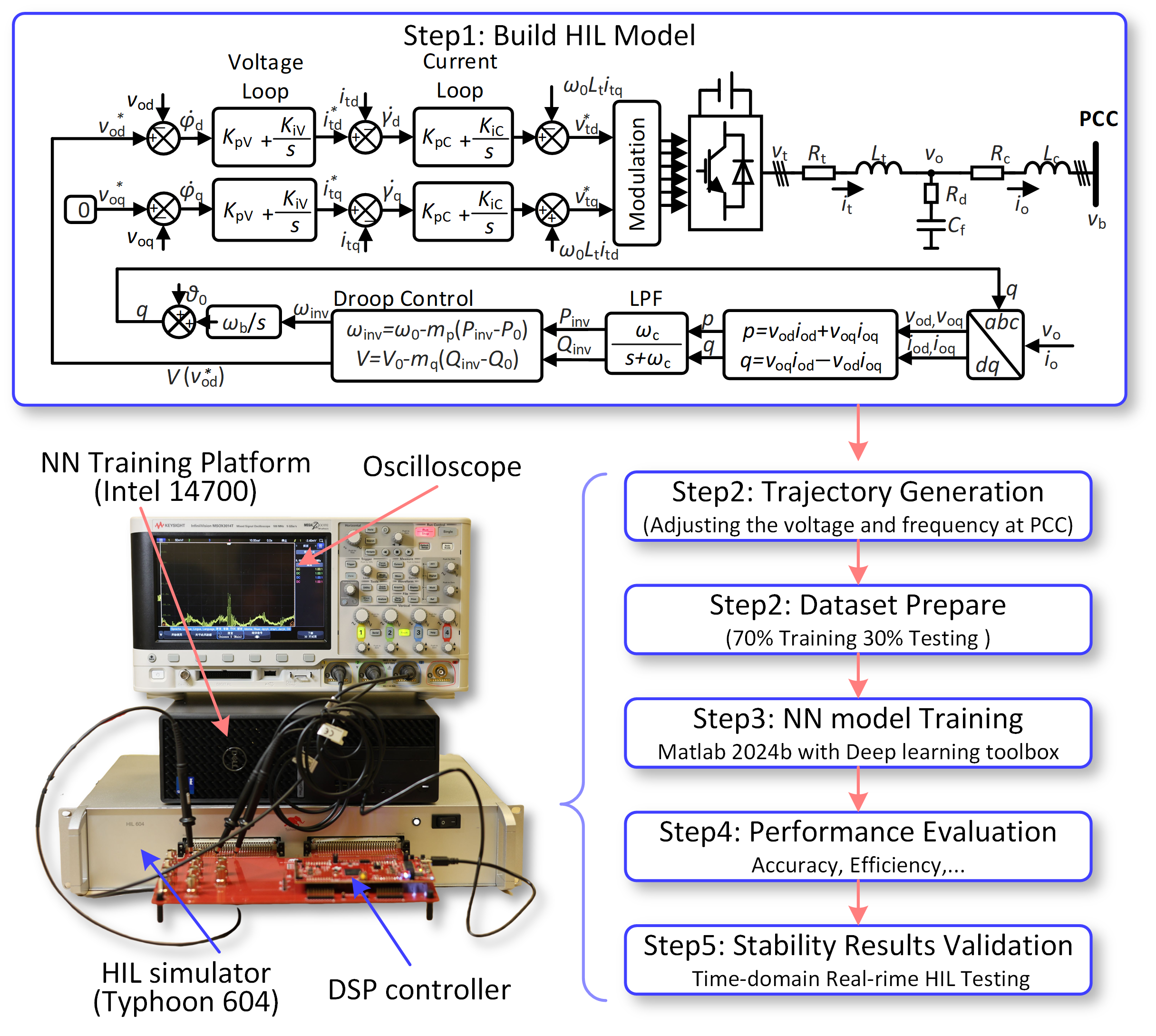} 
    \vspace{-10pt}
    \caption{Hardware platform and evaluation process for the GFM inverter case}
    \label{Ref_Hardware}
\end{figure}

\begin{figure}[t] 
    \centering
    \vspace{-10pt}
    \includegraphics[width=0.46\textwidth]{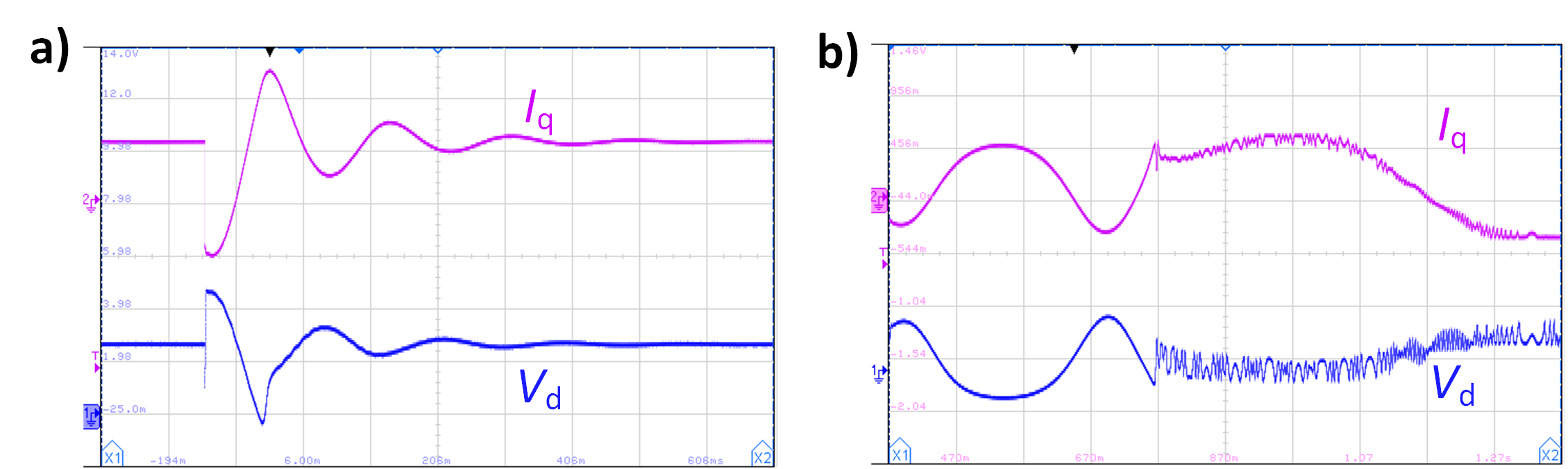} 
    \vspace{-10pt}
    \caption{HIL simulator results  for verification scenario. (a). represents input condition 1. (b). represents input condition 2. }
    \label{Hardware_results}
\end{figure}

\begin{figure}[t] 
    \centering
    \vspace{-10pt}
    \includegraphics[width=0.46\textwidth]{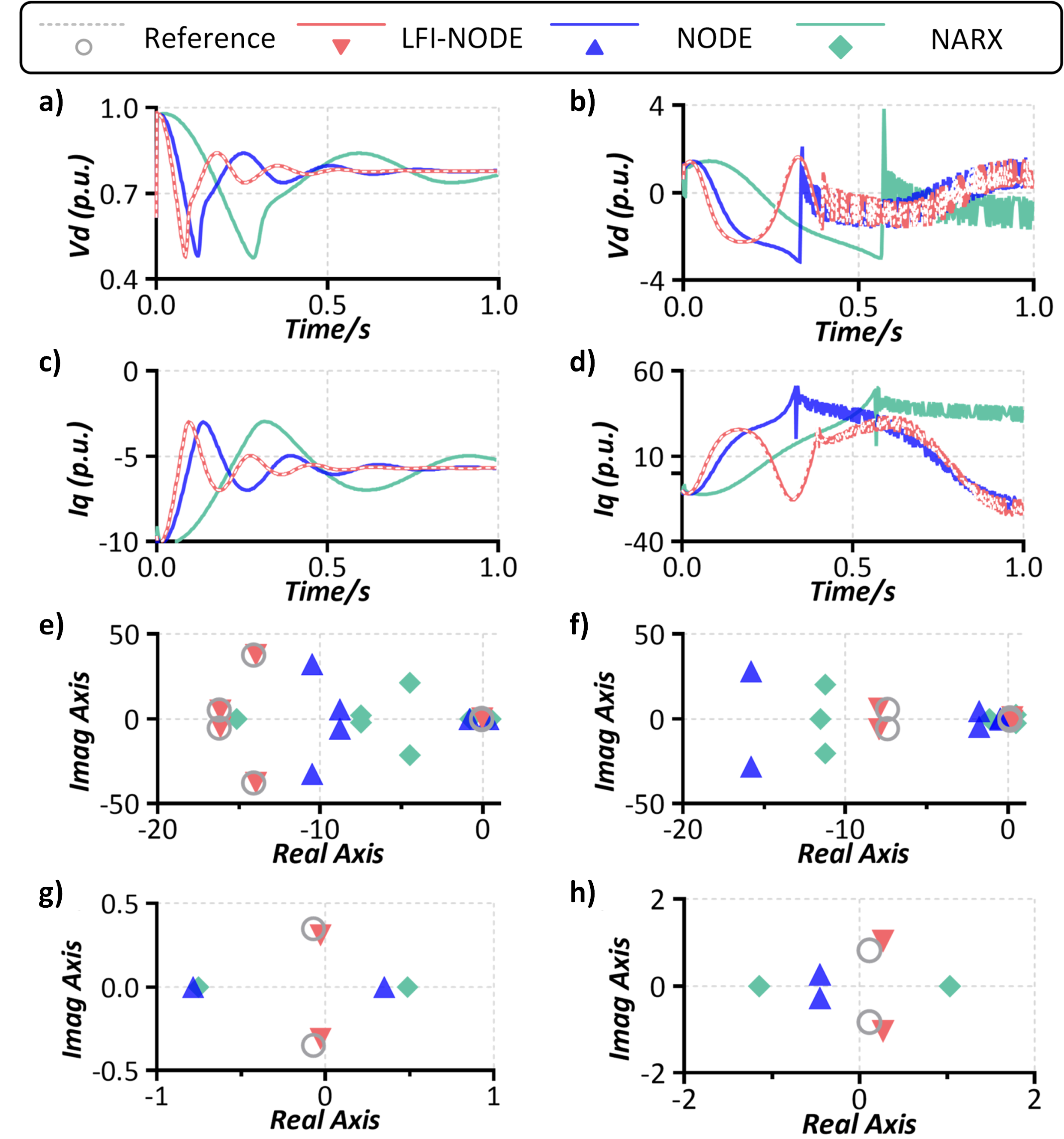} 
    \vspace{-10pt}
    \caption{\revise{Comparison of large-signal behavior and small-signal behavior of the proposed LFI-NODE and NODE, NARX methods under different external conditions. (a), (b) Results of the output voltage Vd. (c), (d) Results of the output current Iq. (e), (f) Eigenvalue results at the stabilization point. (g), (h) Enlarged plot of the eigenvalue near the zero point. The left column (a, c, e, g) represents input condition 1, and the right column (b, d, f, h) represents input condition 2.}}
    \label{Ref_result1}
\end{figure}

\subsection{Case Setup}
This case study targets a grid-forming inverter whose circuit topology, control loops, and parameter values follow the benchmark configuration in \cite{4118327}. As shown in Fig.~\ref{Ref_Hardware}, the inverter is simulated in real time on a Typhoon HIL 604, which both generates training trajectories and evaluates the LFI-NODE model.

Because of limited data in practical settings, only 48 normalized trajectories were generated by sweeping the PCC voltage magnitude and frequency within the safe operating area. \revise{The benchmark model contains 13 state variables, and two external inputs corresponding to the PCC voltage and frequency references. Step perturbations were applied to these two inputs to excite the system dynamics, resulting in a total input dimension of 15 for the neural network.The raw data were sampled at 50~kHz to capture the fast electromagnetic transients dominated by the physical inverter circuits and controllers. Each trajectory lasted 1.0~s, covering the complete response from the disturbance to the new steady state. To reduce computational cost while preserving essential dynamics, the data were preprocessed via zero-phase low-pass filtering for noise suppression and then downsampled, yielding trajectories with 5,000 data points for training.} Training was carried out on an Intel i7-14700 CPU under MATLAB R2024b with the Deep Learning Toolbox.

The proposed LFI-NODE was benchmarked against a conventional NODE and a discrete NARX network. All three models take the 15 inputs and  13 outputs; each adopts a three-layer tanh network with 64–128–128 hidden neurons. LFI-NODE and NODE integrate the dynamics with the AD-enabled solver dlode45 (relative tolerance 
$10^{-7}$, whereas NARX is updated purely in the discrete domain. \revise{The training employs a mini-batch strategy where 40-step sliding windows are sampled from each trajectory. This approach is critical for robustly training the network despite the seemingly small number of 48 trajectories. By treating each trajectory as a dense sequence of thousands of state-derivative pairs, the sliding-window strategy effectively creates a large and diverse training set, which prevents overfitting and ensures the model learns the underlying continuous dynamics. } All other parameters are identical.

\subsection{Performance Evaluation and Comparative Study}

\begin{figure}[t] 
    \centering
    \vspace{-15pt}
    \includegraphics[width=0.48\textwidth]{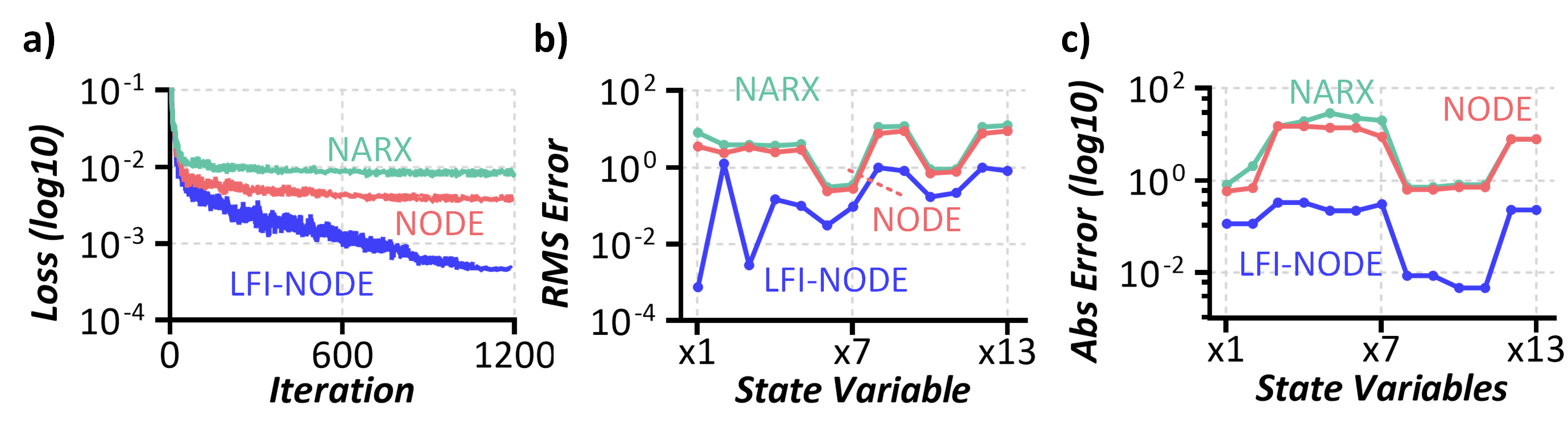} 
    \vspace{-12pt}
    \caption{Quantitative evaluation of different methods in training losses, large dynamic and small dynamic. (a). Training curves. (b). Mean square root error of the time trajectory. (c). Mean absolute error of eigenvalue results.}
    \label{Ref_result2}
\end{figure}

All networks were trained for 1200 iterations and then evaluated on two unseen, normalized input pairs—(0.5,0.9), and (0.8,1.2)—representing stable and unstable cases, respectively. Fig. \ref{Hardware_results} shows the results of the HIL simulator captured from the oscilloscope in the verification scenario, and Fig. \ref{Ref_result1} (Typhoon HIL as ground truth) shows that LFI-NODE matches the reference trajectories and eigenvalues most closely, followed by the standard NODE, whereas NARX exhibits the largest deviation. The superiority of LFI-NODE stems from modeling the continuous ODE directly; the latent-feature guidance further anchors the model around the equilibrium, allowing it to learn the true dynamics with far fewer data.

Fig. \ref{Ref_result2} provides quantitative evidence. Owing to the latent feature term, LFI-NODE converges faster and achieves at least an order-of-magnitude lower training error, as seen in Fig. \ref{Ref_result2}(a). Its large-signal and small-signal prediction errors in Fig. \ref{Ref_result2}(b) and (c) are two orders of magnitude lower than those of the baselines, underscoring the method’s substantial accuracy and practical value.

\subsection{\revise{Robustness Validation under Measurement Noise}}

\revise{To evaluate the robustness of the proposed method against measurement noise, the following study was conducted. Based on the original 48 clean training trajectories, zero-mean Gaussian white noise with different standard deviations was added to the state variables, corresponding to three distinct scenarios: Case 1 ($\sigma_x=1\times 10^{-4}$), Case 2 ($5\times 10^{-4}$), and Case 3 ($2\times 10^{-3}$).  All other training configurations were kept consistent with the baseline setup.}

\begin{figure}[t] 
    \centering
    \vspace{-15pt}
    \includegraphics[width=0.48\textwidth]{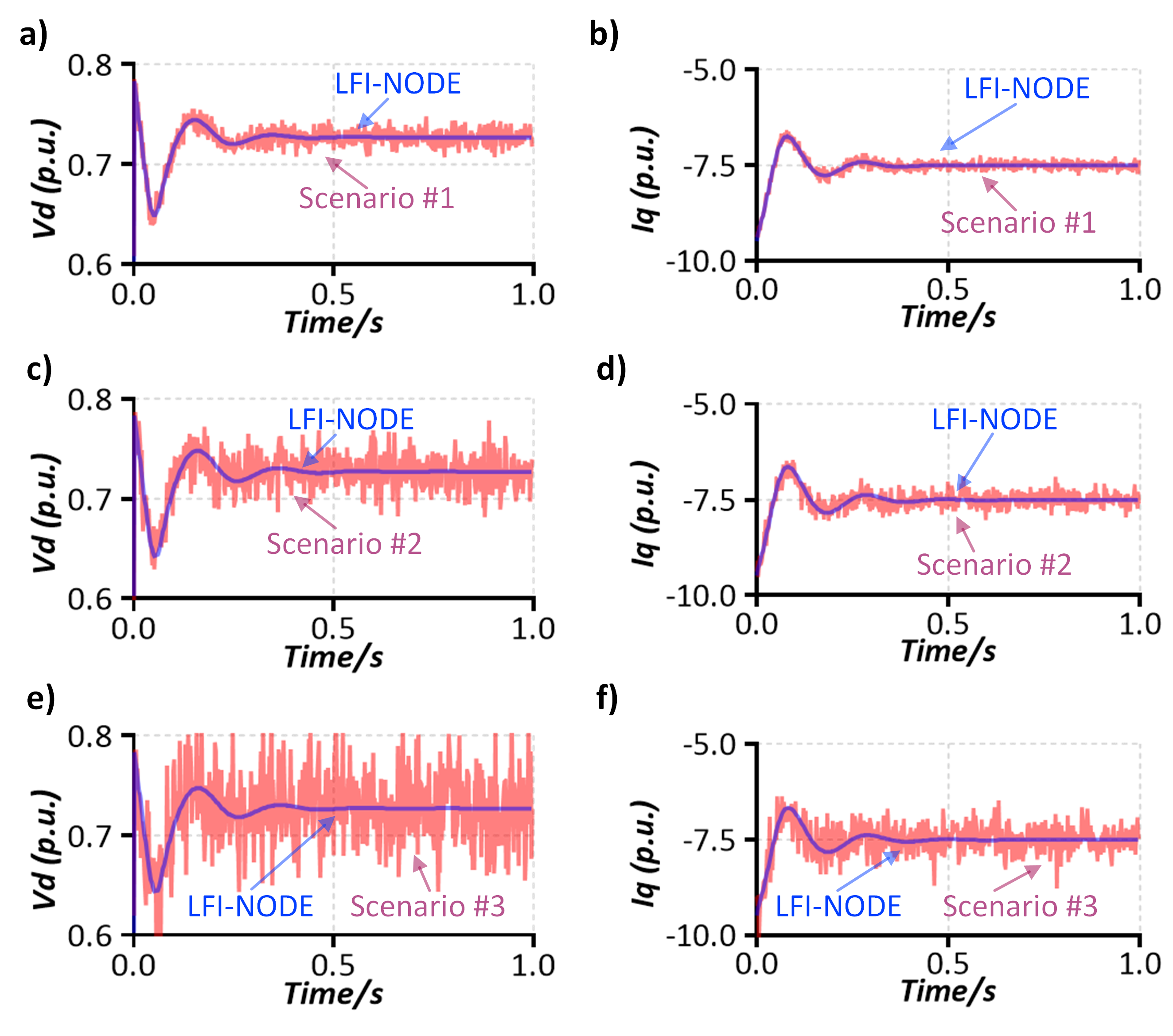} 
    \vspace{-15pt}
    \caption{\revise{Time-domain voltage and current trajectories under different levels of measurement noise: (a)-(b) Case 1, (c)-(d) Case 2, and (e)-(f) Case 3. }}
    \label{Ref_result3}
\end{figure}

\begin{figure}[t] 
    \centering
    \vspace{-12pt}
    \includegraphics[width=0.48\textwidth]{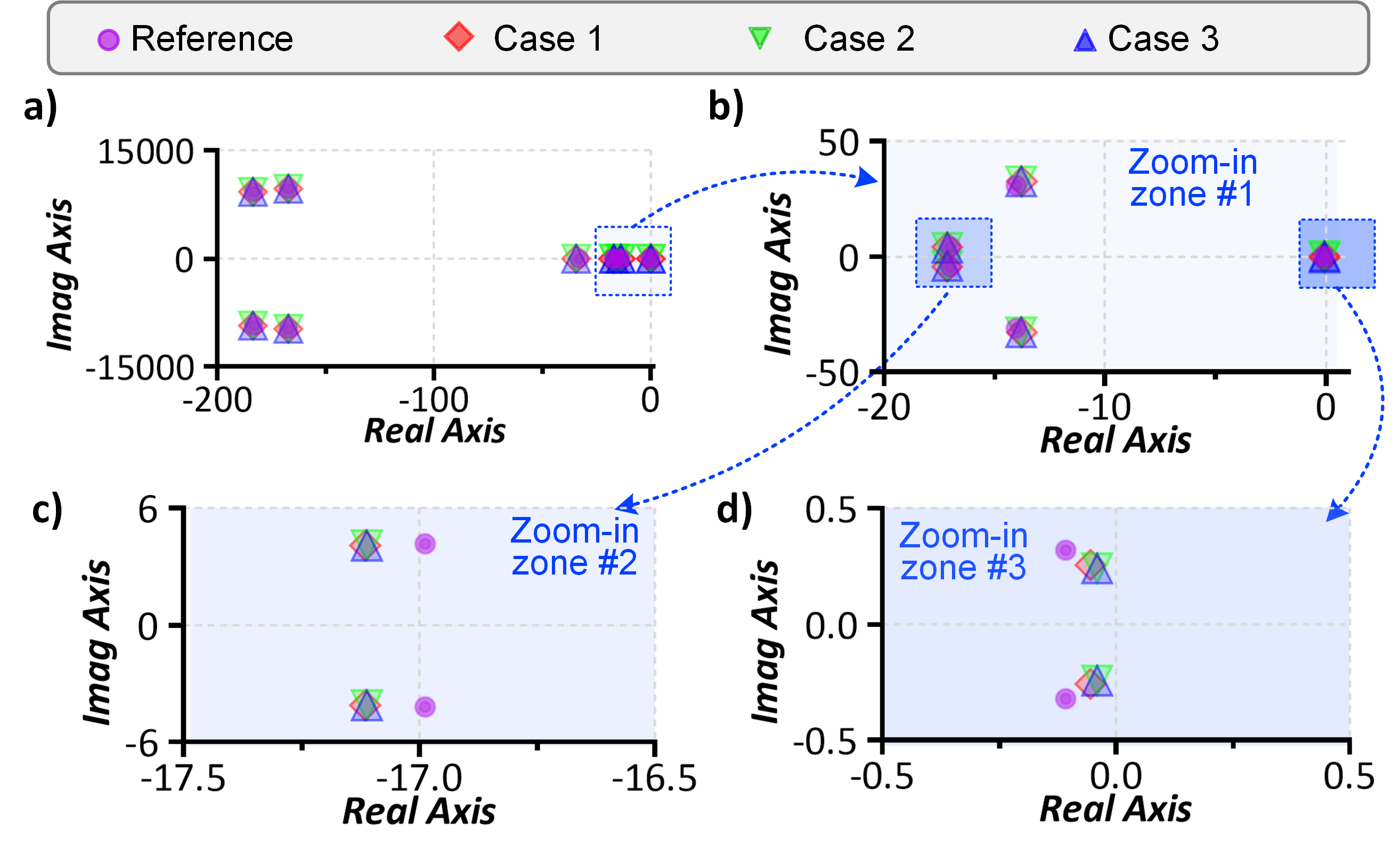} 
    \vspace{-12pt}
    \caption{\revise{Eigenvalue comparison between reference values (purple circles) and the proposed LFI-NODE method under three cases: Case 1 (green diamonds), Case 2 (red squares), and Case 3 (blue triangles). Subfigures (a)-(d) show the eigenvalue distribution on the complex plane with progressive zoom-in views (Zoom-in zones {\#1}-{\#3}) highlighting small deviations. }}
    \label{Ref_result4}
\end{figure}

\revise{The results are shown in Fig.~\ref{Ref_result3} and Fig.~\ref{Ref_result4}. Even at the highest noise level, the predicted large-signal trajectories remain highly consistent with the reference truth, with only minor degradation in smoothness and errors staying within the theoretical bounds. The small-signal eigenvalue predictions also maintain high accuracy, with deviations far smaller than those from baseline models without Jacobian guidance, as previously shown in Fig.~\ref{Ref_result2} (c). These observations confirm that the inclusion of the Jacobian term as a physics-informed regularizer not only improves stability analysis under ideal conditions but also provides strong resilience to noise, in agreement with the error limits derived in Appendix.}

\subsection{Discussions and Future Work}

\revise{Another practical consideration is the method's applicability across different inverter configurations and model types. The LFI-NODE framework is, by design, independent of the inverter's specific configuration. As a black-box identification method, it learns from observed trajectories without prior knowledge of the internal topology or control strategy. The learned model reflects the properties of the training data, capturing either average or switching dynamics accordingly.  For system-level stability analysis, the objective is invariably to identify the inverter's average model. In this work, this was achieved by using data from a HIL platform, which inherently provides the average model dynamics. If data were sourced from physical hardware, a standard pre-processing step involving low-pass filtering would be necessary to extract the desired average dynamic behavior before training.}

\revise{However, for many existing or fully black-box systems, some variables may be infeasible to access. Advanced frameworks such as transformer~\cite{vaswani2023attentionneed}, encoder--decoder~\cite{aitken2021understandingencoderdecoderarchitecturesattend}, and latent mapping~\cite{rubanova2019latentodesirregularlysampledtime}  have been developed to address partially observable systems. A critical issue arises because these frameworks are typically data-hungry, whereas individual power electronic converters are inherently data-scarce, which can lead to poor convergence or overfitting when training with limited data. }

\revise{The LFI-NODE framework is particularly well-suited to address this data-scarcity challenge, opening up a promising direction for future work. A key research avenue is to create a synergy between our data-efficient methodology and these advanced architectures. For example, by integrating the LFI-NODE as the core dynamic model within a Latent ODE structure, it may be possible to robustly train these powerful frameworks on sparse datasets.  
}

\revise{ Furthermore, to enhance robustness under extreme conditions where the operating state deviates significantly from the training data, we  suggest two complementary measures: establishing a trusted operating domain with anomaly detection, triggering alerts when the real-time state moves outside the safe region; and implementing online monitoring and adaptation, where persistent growth in prediction error automatically triggers the collection of new trajectories for fine-tuning or retraining. By integrating LFI-NODE modeling with online adaptation, this approach will not only further improves generalization but also ensures safe and reliable operation in unseen scenarios. }

\section{Conclusion}
This letter has introduced the LFI-NODE modeling method for the stability assessment of black-box grid-tied inverters. By learning the inverter's intrinsic governing ODEs and leveraging latent perturbation features to ensure both large- and small-signal fidelity, this method transcends the traditional dichotomy between time- and frequency-domain data-driven approaches. The demonstrated high accuracy on sparse data confirms that LFI-NODE provides a practical and cost-effective pathway for characterizing the stability of commercial inverters without proprietary information. This physically-informed machine learning paradigm opens new avenues for developing data-efficient, high-fidelity dynamic models for a wide range of complex power electronic systems. \revise{ As a future step, this foundational component-level ODE approach can be scaled to model entire IBR-rich power systems by evolving the learning framework into a Neural Differential-Algebraic Equation (DAE) set to incorporate network algebraic constraints, paving the way for more robust and reliable grid operations.}

\begingroup
\color{black}

\appendix 
 
In this appendix, we analyze the effect of measurement noise on the estimation of the reference Jacobian $J_{\mathrm{ref}}$ in (\ref{ref_Jacobii}), and derive an upper bound for the estimation error.

\subsection{Problem Setup}
Let the true state trajectory be $X(t) \in \mathbb{R}^n$ with its time derivative $\dot{X}(t)$. The true Jacobian at the operating point is denoted by $J_\ast$. In practice, we observe
\begin{equation}
  X_m(t) = X(t) + \eta(t),
\end{equation}
where $\eta(t) \in \mathbb{R}^n$ is additive measurement noise with zero mean and bounded magnitude $\|\eta(t)\|_2 \le \sigma_x$. The derivative is obtained numerically, which introduces noise $\dot{\eta}(t)$ with bound $\|\dot{\eta}(t)\|_2 \le \sigma_{\dot{x}}$.

We collect $N$ samples in a neighborhood of the operating point and construct
\begin{align}
    \Delta X_m &= [X_m(t_1) - X_m(t_0), \dots, X_m(t_N) - X_m(t_0)], \\
    \Delta\dot{X}_m &= [\dot{X}_m(t_1) - \dot{X}_m(t_0), \dots, \dot{X}_m(t_N) - \dot{X}_m(t_0)].
\end{align}

\subsection{Least-Squares Estimation with Noise}
The estimated Jacobian is
\begin{equation}
    J_{\mathrm{ref}} = \Delta\dot{X}_m \,\mathrm{pinv}(\Delta X_m).
\end{equation}
We can write $\Delta X_m = \Delta X + E_x$ and $\Delta\dot{X}_m = \Delta\dot{X} + E_{\dot{x}}$, where $E_x$ and $E_{\dot{x}}$ are the noise matrices.

The least-squares solution without noise is
\begin{equation}
    J_\ast = \Delta\dot{X} \,\mathrm{pinv}(\Delta X).
\end{equation}
The perturbation theory for the pseudoinverse \cite{Wedin1973} gives:
\begin{equation}
    \|J_{\mathrm{ref}} - J_\ast\|_2 \le 
    \kappa(\Delta X) \left( \frac{\|E_{\dot{x}}\|_2}{\|\Delta X\|_2} + \frac{\|E_x\|_2}{\|\Delta X\|_2} \|J_\ast\|_2 \right) + \mathcal{O}(\|E\|^2),
\end{equation}
where $\kappa(\Delta X) = \|\Delta X\|_2 \,\|\mathrm{pinv}(\Delta X)\|_2$ is the condition number of $\Delta X$.

\subsection{Upper Bound under Bounded Noise}
If each noise vector satisfies $\|\eta(t_i)\|_2 \le \sigma_x$ and $\|\dot{\eta}(t_i)\|_2 \le \sigma_{\dot{x}}$, then
\begin{align}
    \|E_x\|_2 &\le \sqrt{N} \,\sigma_x, \\
    \|E_{\dot{x}}\|_2 &\le \sqrt{N} \,\sigma_{\dot{x}}.
\end{align}
Substituting into the perturbation bound yields
\begin{equation}   
    \|J_{\mathrm{ref}} - J_\ast\|_2 \le \kappa(\Delta X) \left( \frac{\sqrt{N}\,\sigma_{\dot{x}}}{\|\Delta X\|_2} + \frac{\sqrt{N}\,\sigma_x}{\|\Delta X\|_2} \|J_\ast\|_2 \right)
\end{equation}
This expression explicitly quantifies the sensitivity of the Jacobian estimate to measurement noise. It shows that robustness improves with a smaller condition number $\kappa(\Delta X)$ and larger trajectory variations $\|\Delta X\|_2$, and benefits from noise suppression in both $X$ and $\dot{X}$.

\endgroup

%

\bibliographystyle{IEEEtran}
\bibliography{Bibliography/BIB_xx-TPEL-xxxx}


\vfill

\end{document}